\documentclass[sigconf,natbib=true]{acmart}

\AtBeginDocument{%
  \providecommand\BibTeX{{%
    \normalfont B\kern-0.5em{\scshape i\kern-0.25em b}\kern-0.8em\TeX}}}

\copyrightyear{2022} 
\acmYear{2022} 
\setcopyright{acmlicensed}\acmConference[SIGIR '22]{Proceedings of the 45th International ACM SIGIR Conference on Research and Development in Information Retrieval}{July 11--15, 2022}{Madrid, Spain}
\acmBooktitle{Proceedings of the 45th International ACM SIGIR Conference on Research and Development in Information Retrieval (SIGIR '22), July 11--15, 2022, Madrid, Spain}
\acmPrice{15.00}
\acmDOI{10.1145/3477495.3531841}
\acmISBN{978-1-4503-8732-3/22/07}

\begin{document}

\title{Modern Baselines for SPARQL Semantic Parsing}


\author{Debayan Banerjee}
\email{debayan.banerjee@uni-hamburg.de}
\affiliation{%
  \institution{Universit\"at Hamburg}
  \city{Hamburg}
  \country{Germany}
}

\author{Pranav Ajit Nair}
\authornote{Both authors contributed equally to this research.}
\email{pranavajitnair.cse18@itbhu.ac.in}
\affiliation{%
  \institution{Indian Institute of Technology (BHU)}
  \city{Varanasi}
  \country{India}
}

\author{Jivat Neet Kaur}
\email{t-kaurjivat@microsoft.com}
\affiliation{%
  \institution{Microsoft Research}
  \city{Bengaluru}
  \country{India}
}
\authornotemark[1]

\author{Ricardo Usbeck}
\email{ricardo.usbeck@uni-hamburg.de}
\affiliation{%
  \institution{Universit\"at Hamburg}
  \city{Hamburg}
  \country{Germany}
}

\author{Chris Biemann}
\email{christian.biemann@uni-hamburg.de}
\affiliation{%
  \institution{Universit\"at Hamburg}
  \city{Hamburg}
  \country{Germany}
}


\begin{abstract}
  In this work, we focus on the task of generating SPARQL queries from natural language questions, which can then be executed on Knowledge Graphs (KGs). We assume that gold entity and relations have been provided, and the remaining task is to arrange them in the right order along with SPARQL vocabulary, and input tokens to produce the correct SPARQL query. Pre-trained Language Models (PLMs) have not been explored in depth on this task so far, so we experiment with BART, T5 and PGNs (Pointer Generator Networks) with BERT embeddings, looking for new baselines in the PLM era for this task, on DBpedia and Wikidata KGs. We show that T5 requires special input tokenisation, but produces state of the art performance on LC-QuAD 1.0 and LC-QuAD 2.0 datasets, and outperforms task-specific models from previous works. Moreover, the methods enable semantic parsing for questions where a part of the input needs to be copied to the output query, thus enabling a new paradigm in KG semantic parsing. Code and data used for this work can be found at \href{https://github.com/debayan/sigir2022-sparqlbaselines}{https://github.com/debayan/sigir2022-sparqlbaselines}.
\end{abstract}


\maketitle

\section{Introduction}
Knowledge Graph Question Answering (KGQA) is the task of finding answers to questions posed in natural language, using triples present in a KG. Typically the following steps are followed in KGQA: 1) Objects of interest in the natural language question (NLQ) are detected and linked to the KG in a step called entity linking (EL).  2) The relation between the objects is discovered and linked to the KG in a step called relation linking (RL). 3) A formal query, usually SPARQL, is formed with the linked entities and relations. The query is executed on the KG to fetch the answer. This step is called Query Building (QB), and is the focus of this paper. We assume that gold entities and relations are made available, and our task is to generate the correct SPARQL query. We experiment with a Pointer Generator Network \cite{see-etal-2017-get} with special vectorisation, and with more recent pre-trained language models such as T5 \cite{t5} and BART \cite{bart}. We choose models that are able to copy tokens from the input text to the output SPARQL query, apart from being able to handle normal category of questions. Questions which require copy operations are found in the dataset LC-QuAD 2.0 \cite{lcq2},
for example: 

\textit{Is it true that an Olympic-size swimming pool's operating temperature is equal to \textbf{22.4} ? }  

which has the corresponding SPARQL over Wikidata KG  \cite{vrandevcic2014wikidata}:

\begin{verbatim}
ASK WHERE 
{ 
  wd:Q2084454 wdt:P5066 ?obj 
  filter(?obj = 22.4) 
}
\end{verbatim}

We found no previous work for SPARQL QB that can handle such questions. We also show that on datasets where copying is not required (LC-QuAD 1.0 \cite{10.1007/978-3-319-68204-4_22} ), the approaches still exhibit a strong performance. 

Our contribution is as follows:

\begin{itemize}
  
  \item We find that with the correct input tokenisation, T5 outperforms all previous works and achieves state of the art performance on LC-QuAD 1.0  over DBpedia \cite{lehmann2015dbpedia}, and LC-QuAD 2.0 over Wikidata.
 \end{itemize}

\section{Related Work}

\citet{10.1007/s10115-017-1100-y} studied several QA systems and concluded that the QB step is generally intertwined with rest of the modules in a pipeline, and hence not evaluated separately. 
Later, with the advent of Frankenstein \cite{frankenstein} and Qanary \cite{qanary}, which are frameworks that allow modular construction of an end-to-end KGQA pipeline, it was 
seen that the QB module is the least researched aspect of the pipeline. Soon after, \citet{10.1145/3178876.3186023} presented a comprehensive survey of individual components of KGQA pipeline and evaluated Sina \cite{SHEKARPOUR201539} and NLIWOD\footnote{\url{https://github.com/semantic-systems/NLIWOD/}} as individual QB components. \\
Several recent works attempt to solve complex query building by constructing intermediate query representations, such as staged query graphs \cite{yih-etal-2015-semantic,hu-etal-2018-state,luo-etal-2018-knowledge, ijcai2020-519, lan-jiang-2020-query} based on $\lambda$-calculus or skeleton structures \cite{Sun_Zhang_Cheng_Qu_2020}. These systems have no built-in method of performing copy operations from input text to output query. The system based on skeleton structure requires manual annotation of queries to corresponding skeletal structures, apart from the final generated query. Some other solutions rely on templates \cite{10.1145/3038912.3052583, ding-etal-2019-leveraging,soru-marx-semantics2017}, which generally have the natural limitation of the query only being limited to the templates chosen beforehand~\cite{DBLP:conf/semco/AthreyaBNU21,DBLP:journals/corr/abs-2103-06752}. In the work of \citet{ding-etal-2019-leveraging}, the templates are discovered from the input training data, however, the implementation of the method of discovery of such templates, structures, and sub-structures in the queries is dataset specific. It is unclear if the same method of discovery can scale to all datasets.\\
Most systems that work over query graphs and use $\lambda$-calculus start with the assumption that entities are already linked, and the relations must still be found. However, with the availability of joint entity and relation linkers such as EARL \cite{earl} and Falcon \cite{falcon} , we can develop semantic parsers which focus only on query building and receive both entities and relations pre-linked.\\
For non-KG semantic parsing, PLMs have been evaluated recently with a focus on compositional generalisation \cite{shaw}. For KG semantic parsing, the Compositional Freebase Questions (CFQ) \cite{cfq} dataset that is based on the Freebase KG \cite{freebase}, has reported results with non-pre-trained Transformer-based models, while subsequent works \cite{herzig} \cite{furrer} over the same dataset have experimented with PLMs. This line of work may be considered closest to ours, however, their focus is on compositional generalisation, while our focus is on the ability to faithfully copy input tokens to the produced query. CFQ has no questions that require such copy operations. Moreover, since Freebase is no longer an active project, our focus is on DBPedia and Wikidata. By exploring two new datasets and corresponding KGs, we expand the scope of research. \\
Some recent work on Complex Temporal Question Answering \cite{ctqa} handled questions that require extraction of time stamps from the input question. They use task-specific temporal extraction tools such as SUTime \cite{sutime} and HeidelTime \cite{heideltime} for this purpose, however in the end, they do not form a SPARQL query, and instead attempt to find the correct answer directly through an entity re-ranking approach. Our efforts are instead on finding such flexible single models which can perform such extraction tasks along with the building of SPARQl queries.  \\
The models we experiment with pose no structural or template-based constraints neither at an intermediate stage nor during decoding, and are able to produce any possible query structure since we generate the final query token by token. They require the input question, the linked entities and relations, and a fixed vocabulary of all possible SPARQL tokens. This allows them to operate on all possible datasets and KGs.


\begin{figure*}[htb!]
  \centering
  \begin{minipage}[b]{1.0\textwidth}
    \centering
    \includegraphics[width=0.9\textwidth]{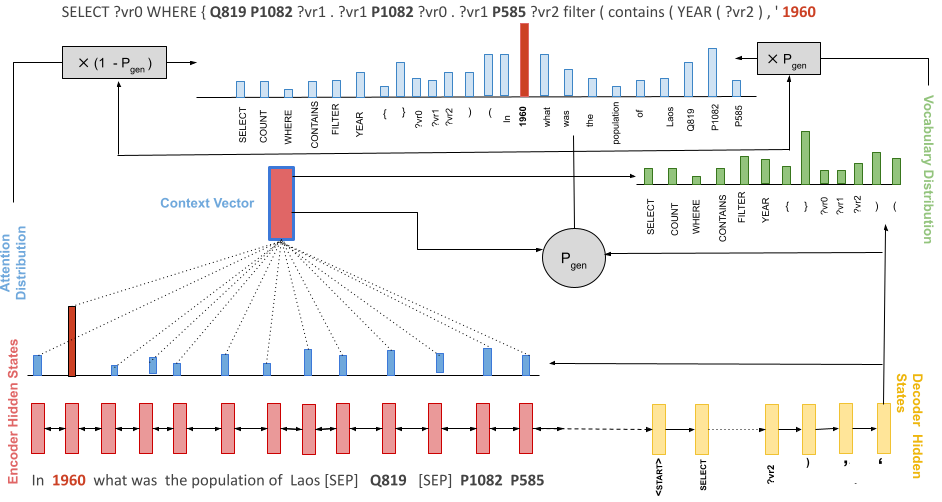}
  \end{minipage}
  \hfill
  \caption{PGN-based QB model. At the current time step, the model is decoding the symbol after the single quote character ('). It considers the scores over the vocabulary and the attention weights over the input text to obtain a final probability distribution, from which it makes the prediction of choosing \textit{1960} as the next token.}
\label{figurepgn}
\end{figure*}

\section{Models}
\subsection{T5}
T5, or Text-to-Text Transfer Transformer, is a Transformer \cite{transformers}  based encoder-decoder architecture that uses a text-to-text approach. Every task, including translation, question answering, and classification, is cast as feeding the model text as input and training it to generate some target text. This allows for the use of the same model, loss function, hyper-parameters, etc. across a diverse set of tasks. We experiment with two variants of the T5 model: \texttt{T5-small} with 60 million parameters and \texttt{T5-base} with 220 million parameters. The uncompressed C4 corpus used to pre-train the models is 750 GB in size.

\subsubsection{Input Tokenisation}
\label{t5tokenisation}
Let the natural language input question be denoted by $Q = [w1,w2 ... wn]$ where 
$w$ denotes the words in the question. The linked entities be denoted by $E = [E1,E2...En]$ and the relations by $R = [R1,R2....Rn]$. For each entity and relation we fetch the corresponding label from the respective KG such that $E_{{lab}_1}$ denotes the label for $E_1$, and similarly for relations. Let the SPARQL vocabulary be donated by $V$. The input string to the model is formed as follows: \\

$ w_1 w_2 ... w_n  [SEP] E_1 E_{{lab}_1} .... E_n E_{{lab}_n} [SEP] R_1  R_{{lab}_1} ... R_n  R_{{lab}_n} $ \\

 where each word is separated from the other word by a space.
We denote the entities by their respective KG IRIs. For DBpedia the entities take the form, e.g. " \texttt{http://dbpedia.org/resource/Dolley\_
Madison}" while the relations look like " \texttt{http://dbpedia.org/onto
logy/spouse}". For Wikidata a typical entity looks like \texttt{wd:Q76} (Barack Obama) while a typical relation looks like \texttt{wdt:P31} (instance of). The prefixes \texttt{wd:}, \texttt{wdt:} and several others are expanded according to the list available online\footnote{\url{https://www.wikidata.org/wiki/EntitySchema:E49}}. We shuffle the order of entities and relations randomly before passing it to the model. \\
When the input string is passed to T5 as it is, the accuracy is close to zero, as the \texttt{AutoTokenizer} splits the URLs as it sees fit, and it later fails to concatenate the fragments properly at the output. To work around this problem for DBpedia, we made use of the 100 sentinel tokens that T5 provides. The sentinel tokens were originally used during pre-training objective to denote masked tokens, but in our case, we make use of them to represent the prefixes, e.g, \texttt{http://dbpedia.org/ontology/} as a specific sentinel token such as \texttt{<extra\_id\_2>}. Additionally, we represent each of the items in the SPARQL vocabulary $V$ as a sentinel token with a specific ID for each keyword. We do the same for prefixes in Wikidata for LC-QuAD 2.0.\\

\subsubsection{Output}
The output is composed of tokens from $Q, V, E$ and $R$. In the case of LC-QuAD 1.0, since there are no questions that require copying of tokens from Q, the output is composed of tokens from $V,E$ and $R$. We would like to point out that since our input contains the entities and relations, the model always has to perform a form of "copying" even to decide where in the output SPARQL query to place the entities and relations. In the case of questions from LC-QuAD 2.0, in some cases this copying mechanism must also decide which input token from $Q$ to copy and to which position in the output. The output for $E$ and $R$ are not produced verbatim, and instead they are split as a special token prefix, and then the entity or relation ID. A post-processing step is required to resolve these sentinel token IDs into the original prefixes and combine them with the adjacent IDs. For example, \texttt{wdt:P31} is produced in the output as \texttt{<extra\_id\_3> P31}.\\

\subsection{BART}
The Bidirectional and Auto-Regressive Transformer or BART is a Transformer that combines the Bidirectional Encoder with an Autoregressive decoder into one Seq2Seq model. We experiment with the \texttt{BART-base} model consisting of 139 million trainable parameters. This model was pre-trained on 160 GB of data resulting from the combination of four corpuses, namely, Bookcorpus \cite{bookcorpus}, CC-News \cite{ccnews}, OpenWebText \cite{openwebtext} and Stories \cite{stories}.

\subsubsection{Input Tokenisation}

We form the input string for BART in the same manner as discussed in Section \ref{t5tokenisation} for T5. However there is no concept of sentinel tokens in BART. Hence to handle special tokens, we instead add them to the \texttt{BartTokenizer} using the \texttt{add\_tokens} function and appropriately resize the token embedding space using the \texttt{resize\_token\_embeddings} function.\footnote{\url{https://huggingface.co/docs/transformers/model_doc/bart}}

\subsection{Pointer Generator Network}
A PGN is a seq2seq network with an encoder and a decoder block. We experiment exclusively with LSTM-based \cite{lstm} PGNs. While PGNs can generate output tokens from a given vocabulary via softmax selection, they are also able to copy tokens from the input text to the output by making use of attentions weights of the decoder. A significant difference between T5, BART and PGN is that for PGNs no pre-training on corpus is performed. The model consists of 53 million trainable parameters which is comparable to \texttt{T5-small}'s 60 million parameters size.  

PGNs have typically been employed for abstract summarisation tasks but recently \citet{Rongali_2020} used PGNs to perform SQL semantic parsing.  Figure \ref{figurepgn} depicts this architecture parsing an example query. 

\begin{table}[htb!]
  \centering
  
  \begin{tabular}{|c|c|c|}
    \hline
      & F1 & F1 \\
    \hline
      & DBpedia 16.04 & Wikidata \\
    \hline
      & LC-QuAD 1.0 & LC-QuAD 2.0 \\
    \hline
     Sina \cite{SHEKARPOUR201539}             & 0.24 & - \\
     NLIWOD     & 0.48 & - \\
     SQG \cite{10.1007/978-3-319-93417-4_46}  & 0.75 & - \\
     CompQA \cite{luo-etal-2018-knowledge}    & 0.77 & - \\
     SubQG \cite{ding-etal-2019-leveraging}   & 0.85 & - \\
     AQG-net \cite{ijcai2020-519}                    &  -   & 0.45  \\
     Multi-hop QGG  \cite{lan-jiang-2020-query}      &  -   & 0.53  \\
     CLC \cite{zou2021chinese}                       &  -   & 0.59\\
     \hline
    PGN-BERT                                  & 0.67    & 0.77 \\
    PGN-BERT-BERT                             &  0.88   & 0.86 \\
  
    BART                                      &   0.84  & 0.64 \\
    T5-Small                                  &   0.90  & \textbf{0.92}   \\
    T5-Base                                   &   \textbf{0.91}  & 0.91 \\
    \hline
  \end{tabular}
  
  \caption{Results for query generation with gold entities and relations. Best results are in bold.}
  \label{table:results}
\end{table}

\vspace*{-\baselineskip}

\begin{figure}[htb!]
\centering
\includegraphics[width=\columnwidth, height=3cm]{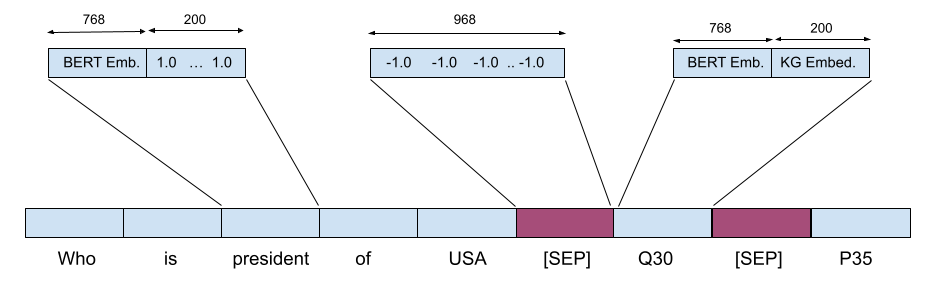}
\caption{Input vector for PGN-BERT.}
\label{fig_input}
\end{figure}

\subsubsection{Input Tokenisation}
\label{sec: input_vec}

 We construct our input vector (Figure \ref{fig_input}) by concatenating the question tokens with gold entities and relations. We add a separator token \texttt{[SEP]} between question tokens and entity candidates, as well as between entity candidates and relation candidates to help the model understand the segments of the input tokens. Each token is represented by a 968-dimensional vector, where the first 768 dimensions are the BERT contextual embeddings of the question token, or the entity label or relation label, as the case may be. The next 200 dimensions are reserved for the KG embeddings. For question tokens, KG embeddings do not apply, so we fill the 200 dimensions with a sequence of \texttt{1.0}. The \texttt{[SEP]} token is represented by a vector full of \texttt{-1.0}. The linked entities and relations carry the respective KG embeddings in the last 200 dimensions.

\subsubsection{KG Embeddings}

For both datasets, we make us of TransE embeddings, due to its popularity and ease of availability \cite{10.5555/2999792.2999923}. For DBpedia, We train TransE embeddings ourselves using PyTorch-BigGraph \cite{pbg}. For the training, we proceed to 30 epochs and use the cosine dot operator as the comparator function. For Wikidata we made use of readily available TransE embeddings made available by Pytorch-BigGraph\footnote{\url{https://github.com/facebookresearch/PyTorch-BigGraph}} on their webpage. 

\subsubsection{Re-Ranker}

We found that the beams produced by the PGN often contained the right query but not in the top position. To improve the ranking of queries further, we fine-tuned a pre-trained BERT-based classifier (\texttt{distil-bert-base-uncased}) on the output of the PGN. We take the top-10 beam outputs of the PGN on the val sets and query the KG. We save the output of all the queries which produce a valid response from the KG. To train the BERT-based classifier, we form the input string by concatenating the question tokens, the SPARQL query, and the KG response. For supervision, we provide binary labels 0 and 1 for right and wrong query. Once trained, we use the logit values of the penultimate layer of the model for re-ranking the queries produced by PGN on the test set.

\section{Datasets}

\begin{itemize}

\item\textbf{LC-QuAD 1.0}  contains 5,000 questions that are answerable on DBpedia 2016-04. The dataset originally contains a 4:1 train-test split. However, several competing systems in Table 1 instead use a smaller split of 3,253 questions. We adopt the same split and follow \citet{ding-etal-2019-leveraging} in performing 5-fold cross-validation with 70:10:20 split for train, dev, and test, respectively. There are no questions that require copying of input tokens to output query in this dataset.
\item\textbf{LC-QuAD 2.0}  is based on Wikidata and consists of a mixture of simple and complex questions that were verbalized by human workers on Amazon Mechanical Turk. It is a large and varied dataset comprising 24,180 train questions and 6,046 test questions. In addition, to the best of our knowledge, this is the sole dataset that incorporates aspects of hyper-relational \cite{StarE} structure of recent Wikidata versions. Approximately 16\% of questions in the dataset require copying of input tokens to the output query. 

\end{itemize}

\section{Evaluation}
\label{evaluation}

In Table \ref{table:results}, the results of competing systems for LC-QuAD 1.0 are shown as reported by \citet{ding-etal-2019-leveraging}. The results for competing systems for LC-QuAD 2.0 are as reported by \citet{zou2021chinese}, also found on the KGQA leaderboard ~\cite{DBLP:journals/corr/abs-2201-08174}\footnote{\url{https://kgqa.github.io/leaderboard/}}.  For evaluating T5, BART and PGNs on both datasets, we take the top-10 beams outputs of the models and query the KG in ranked serial order. The first query to return a non-empty response is considered the output of the model. We match the KG responses of the generated query to the gold query and mark it a match if they are identical, and no match if they are not.

In the case of LC-QuAD 2.0, we found a large number of queries no longer answerable on the current version of Wikidata. The original dump used to create the dataset is no longer available online\footnote{\url{https://databus.dbpedia.org/dbpedia/wikidata/debug/2020.07.01}}. As a result, we setup an endpoint with the dump dated 13 October 2021\footnote{\url{https://dumps.wikimedia.org/wikidatawiki/entities/}} of Wikidata and filtered the test set of 6046 questions down to 4211 questions for which the gold query produced a valid response. We loaded the triples to a Virtuoso Open Source Triple Store and used it as our endpoint. \citet{zou2021chinese} resort to a similar pre-filtration step for the dataset in their work.

In Table \ref{table:results}, \texttt{PGN-BERT} refers to the setting where we feed the PGN with BERT embeddings. We apply no re-ranker to the output beams, and consider the first query in the ranked beams to produce a valid response from the KG as the output of the system. On the other hand, \texttt{PGN-BERT-BERT} refer to the setting where the beams are further re-ranked with the BERT-based re-ranker described previously.

\section{Discussion}
\label{discussion}

T5 not only outperforms all previous works on both datasets, but also BART and PGN that we experiment with. It is important to handle input tokenisation of prefixes through sentinel IDs for T5, or else the model produces zero accuracy. In the case of PGN, we also performed a test without input KG embeddings, where we saw a 20\% point drop in accuracy. In spite of the addition of KG embeddings, it fails to get an edge over T5, which operates solely based on entity and relation labels. It can hence be concluded that the large amount of pre-training for T5 outranks the KG embedding advantage.

In the case of BART versus T5, we observed that the copy actions of BART were not as clean as T5. For example, in LC-QuAD 2.0, BART produced an F1 of 0.3 for questions that require token copy to the output, while T5 produced an F1 of 0.8. We believe this is due to the fact that instead of using sentinel IDs, which are inherently a part of the T5 model from its pre-training objective, we had to resort to adding special tokens to the \texttt{BartTokenizer}, which does not produce the same copying performance. 

\texttt{T5-small} and \texttt{T5-base} produce similar performance, which suggests that the extra number of parameters in the base model remains unused for our given task. For LC-QuAD 2.0, none of the previous works has the ability to copy tokens from input to the output, so our approaches outperform them by far. 
In terms of PGN performance, on using the BERT re-ranker, we see gains in the range of 10-20 \% points, which shows that PGN does a poor job of producing the right query towards at the top. When comparing \texttt{T5-small} performance with PGN, although they have similar parameter count, the pre-training on 750 GB of corpus seems to be the crucial factor in \texttt{T5-small}'s favour.

\section{Error Analysis}

We randomly sampled 100 cases of erroneous outputs for PGN, T5-Small and BART, as shown in Table \ref{table:erroranalysis}. The most important difference in the errors produced by T5/BART versus PGN is what we call "copy morphing". T5/BART produce their output from an open ended sub-word-based vocabulary, and hence sometimes corrupt the item being copied. At times, they produce wrong entity IDs, e.g., in the case of DBpedia, \texttt{Barack-Obama} instead of \texttt{Barack\_Obama}, and in some other cases, they hallucinate entity IDs that are not part of the linked entities or relations in input. We also encountered cases of morphing relation IDs to semantically similar words, like \texttt{notableWork} changing to \texttt{notabilityWork}, or unexpected capitalisation, e.g., \texttt{Artist} instead of \texttt{artist}. Although PGN's overall accuracy is lower, they do not exhibit this issue, since the copying takes place from an expanded vocabulary which is limited, consisting of SPARQL tokens and the input tokens. We also found that copy morphing in T5 takes place only for DBpedia where the entity and relation IDs are similar to the dictionary labels. For Wikidata, where the IDs are numerical, morphing does not occur.

We categorise the other kind of errors under the following heads: Triple Flip, e.g., \texttt{<o p s>} instead of \texttt{<s p o>}, Wrong Variables, e.g., \texttt{?var0} instead of \texttt{?var1}, Wrong Intent, e.g., \texttt{ASK} instead of \texttt{SELECT}, Copy Errors, where the wrong token is copied, and Syntax Errors, where the generated query is malformed and an invalid SPARQL query. We observe that BART has poor performance in copying, while other models have most errors in triple flips and wrong variables.


\begin{table}[h]
  \centering
  \begin{tabular}{|c|c|c|c|c|c|c|}
  \hline
    & \multicolumn{2}{|c|}{PGN} & \multicolumn{2}{|c|}{BART} & \multicolumn{2}{|c|}{T5-Small} \\
\hline
    &LCQ1  &LCQ2 &LCQ1  &LCQ2&LCQ1&LCQ2\\ 
\hline
      Triple Flip   &56    &54   & 22     & 30  &66 & 60 \\
\hline       
      Wrong Var     &78    &63   & 18    & -  &36 & 8 \\
\hline
       Wrong Intent  &22    &15   &  -    & 10 &4 & 26 \\
    
\hline
      Copy Error    & -    & 14  & 22     & 40 & - & 18 \\
\hline

      Copy Morph      & -    & -   & 60     & - &26 & - \\
\hline
      Syntax Errors   & -    & -   & 16     & 40 & - & - \\
      
     \hline
  \end{tabular}

  \caption{Error breakdown for randomly sampled 100 errors}
  
  \label{table:erroranalysis}
\end{table}

\section{Conclusion and Future Work}

In this work, we establish new baselines using PLMs for the KG semantic parsing task for two datasets. Our results surpass the results of earlier baselines and lay grounds for future research. We make use of readily available and popular models, with minimal changes to input vectorisation and tokenisation, to report state of the art results on SPARQL semantic parsing and query building tasks over DBpedia and Wikidata. We show that these methods are flexible and able to handle a variety of questions with no fixation on templates or query graphs. The direction of future semantic parsing work for SPARQL should lay more emphasis on pre-training, since currently it is more focused on producing custom model architectures for the tasks.

As future work, we would like to explore the ability of these models in disambiguation tasks, where the input consists of entity and relation candidates, instead of linked entities and relations.

\section{Acknowledgments}

This research was partially funded by the German Federal Ministry of Education and Research (BMBF) as part of the INSTANT project, ID 02L18A111.

\bibliographystyle{ACM-Reference-Format}
\bibliography{sample-base}

\end{document}